# Optical Field Enhancement in Nanoscale Slot Waveguides of Hyperbolic Metamaterials

Yingran He[1,2], Sailing He[2], and Xiaodong Yang[1,*]

[1] Department of Mechanical and Aerospace Engineering, Missouri University of Science and Technology, Rolla, MO 65409, USA

[2] Centre for Optical and Electromagnetic Research, Zhejiang Provincial Key Laboratory for Sensing Technologies, Zhejiang University, Hangzhou 310058, China

Corresponding author: [*]yangxia@mst.edu

**Abstract:** Nanoscale slot waveguides of hyperbolic metamaterials are proposed and demonstrated for achieving large optical field enhancement. The dependence of the enhanced electric field within the air slot on waveguide mode coupling and permittivity tensors of hyperbolic metamaterials is analyzed both numerically and analytically. Optical intensity in the metamaterial slot waveguide can be more than 25 times stronger than that in a conventional silicon slot waveguide, due to tight optical mode confinement enabled by the ultrahigh refractive indices supported in hyperbolic metamaterials. The electric field enhancement effects are also verified with the realistic metal-dielectric multilayer waveguide structure.

It was proposed by Almeida *et al.* that considerable electric field enhancement will occur at the low-index slot region of coupled high-index waveguides, as a result of the normal electric displacement continuity at the high-index-contrast interface implemented by Maxwell's equations [1]. Silicon slot waveguides are attractive, as silicon can provide a high refractive index ($n_{si}$ = 3.48 at 1.55 µm) and thus a large field enhancement [2, 3]. For further boost of the optical field inside the low-index slot region, a material with a higher refractive index is preferable. Although high indices of refraction at optical frequencies are not available in natural materials, metamaterials with artificially engineered unit cells can be carefully designed to achieve this goal [4, 5]. Particularly, hyperbolic metamaterials with extreme anisotropy, in which not all the principal components of the permittivity tensor have the same sign, can support large wave vectors and therefore ultrahigh effective refractive indices, due to the unique spatial dispersion [6-10].

In this work, we will present a new type of slot waveguides made of hyperbolic metamaterials with ultrahigh refractive indices, where the optical field within the slot region can be significantly enhanced. The dependences of the electric field enhancement on gap sizes between two coupled waveguides and permittivity tensors of hyperbolic metamaterials are systematically studied with both numerical simulation and theoretical analysis. It is demonstrated that the maximum optical intensity in the slot waveguides of hyperbolic metamaterial can be more than 25 times stronger than that in a silicon slot waveguide, due to the ultrahigh effective refractive indices in hyperbolic metamaterials. The proposed metamaterial slot waveguides with extremely tight photon confinement will be of great importance in the enhancement of light-matter interactions, such as optical trapping [3], nonlinear optics [11], optomechanics [12], and quantum electrodynamics [13].

Fig. 1(a) shows the schematic of the proposed metamaterial slot waveguides. Two identical waveguides with square cross sections (both width $w$ and height $h$ are equal to 80 nm) are closely placed with a nanoscale gap $g$ along the $y$ direction. In each waveguide, the hyperbolic metamaterial is constructed with alternative thin layers of silver (Ag) and germanium (Ge) along the $y$ direction. The multilayer metamaterial can be treated as a homogeneous effective medium and the principle components of the anisotropic permittivity tensor can be determined from the effective medium theory (EMT), [14, 15]

$$\varepsilon_x = \varepsilon_z = f_m \varepsilon_m + (1 - f_m) \varepsilon_d$$
$$\varepsilon_y = \frac{\varepsilon_m \varepsilon_d}{f_m \varepsilon_d + (1 - f_m) \varepsilon_m} \quad (1)$$

where $f_m$ is the volume filling ratio of silver, $\varepsilon_d$ and $\varepsilon_m$ are the permittivity corresponding to germanium and silver, respectively. The permittivity of germanium is $\varepsilon_d = 16$, and the optical properties of silver are described by the Drude model $\varepsilon_m(\omega) = \varepsilon_\infty - \omega_p^2/(\omega^2 + i\omega\gamma)$, with a background dielectric constant $\varepsilon_\infty$ = 5, plasma frequency $\omega_p$ = 1.38×10$^{16}$ rad/s and collision frequency $\gamma$ = 5.07×10$^{13}$ rad/s. At the telecom wavelength 1.55 μm, the permittivity tensors of hyperbolic metamaterial are $\varepsilon_y$ = 29.2 + 0.12i, $\varepsilon_x = \varepsilon_z$ = -39.8 + 2.1i for $f_m$ = 0.4, and $\varepsilon_y$ = 76.3 + 1.4i, $\varepsilon_x = \varepsilon_z$ = -81.7 + 3.6i for $f_m$ = 0.7, respectively.

Fig. 2(a) shows the mode profiles of the metamaterial slot waveguides for the filling ratio $f_m$ = 0.4, and the gap size $g$ = 10 nm, calculated by the finite-element method (FEM) with the software package COMSOL. A considerable enhancement of electric field $E_y$ is observed at the slot region as a result of the abrupt jump of normal permittivity $\varepsilon_y$ at the interface of the metamaterial waveguide and the air slot. Due to the strong electric field enhancement, a large fraction of optical power flow is localized within the slot region, as clearly seen from the profile of optical power flow density $S_z$ in Fig. 2(a). It is noticed that the evanescent optical field leakage into the air surrounding the metamaterial waveguides is quite weak compared to the conventional dielectric waveguides, due to the ultrahigh refractive indices supported in metamaterial waveguides. The nanoscale gap size $g$ is expected to play a critical role in the waveguide coupling and therefore the optical field enhancement in the slot region. Fig. 2(b) gives a comparison of $E_y$ field distributions for two gap sizes, $g$ = 10 nm and $g$ = 5 nm. It is clear that a smaller gap size will lead to a stronger optical coupling between metamaterial waveguides and consequently a larger enhancement factor of the electric field.

In order to understand the mechanism of electric field enhancement inside metamaterial slot waveguides, theoretical analysis is conducted based on the 2D coupled slab waveguides shown in Fig. 1(b), which is approximated from the 3D coupled metamaterial waveguides. Here the 2D slab waveguide approximation is valid, due to the fact that the mode profiles in 3D coupled waveguides plotted in Fig. 2(a) show negligible dependence on the $x$ coordinate, so that the wave vector $k_x \approx 0$. Assuming the optical field components of the slot waveguides have the form of exp(i$\beta$z-i$\omega$t), the coupled electric field can be expressed as follows,

$$E_y = E_0 \begin{cases} \cos\left(-k_y \dfrac{h}{2}+\varphi\right)\dfrac{\cosh(\gamma y)}{\cosh(\gamma \dfrac{g}{2})}, & 0<|y|<\dfrac{g}{2} \\ \dfrac{1}{\varepsilon_y}\cos\left[k_y\left(|y|-\dfrac{h+g}{2}\right)+\varphi\right], & \dfrac{g}{2}<|y|<\dfrac{g}{2}+h \\ \cos\left(k_y \dfrac{h}{2}+\varphi\right)\exp\left[-\gamma(|y|-\dfrac{g}{2}-h)\right], & |y|>\dfrac{g}{2}+h \end{cases} \qquad (2)$$

where $\beta$ is the propagation constant of the waveguide mode, $\omega$ is the angular frequency corresponding to $\lambda_0 = 1.55$ μm, $\varphi$ is the phase shift of optical field at the middle of each waveguide due to the waveguide coupling effect. The wave vector inside metamaterial $k_y$ and the field decay rate in air $\gamma$ are related to $\beta$ through the following dispersion relations for hyperbolic metamaterial and air, respectively,

$$\dfrac{\beta^2}{\varepsilon_y}+\dfrac{k_y^2}{\varepsilon_z}=k_0^2$$
$$\beta^2-\gamma^2=k_0^2 \qquad (3)$$

where $k_0$ is the vacuum wave vector corresponding to $\lambda_0 = 1.55$ μm. By applying the continuity conditions of tangential field components $E_z$ and $H_x$ at the metamaterial-air interface, we obtain the following characteristic equations for metamaterial slot waveguides,

$$\tan\left(-k_y\dfrac{h}{2}+\varphi\right)=-\dfrac{\gamma\varepsilon_z}{k_y}\tanh(\dfrac{\gamma g}{2})$$
$$\tan\left(k_y\dfrac{h}{2}+\varphi\right)=\dfrac{\gamma\varepsilon_z}{k_y} \qquad (4)$$

By solving the above equations, all the optical field properties including the propagation constant $\beta$ can be obtained analytically, which will provide a physical interpretation for the mechanism of optical field enhancement in hyperbolic metamaterial slot waveguides.

Fig. 3(a) presents the effective refractive index along the propagation direction $n_{\text{eff},z} \equiv \beta/k_0$ of the metamaterial slot waveguides as a function of gap size $g$ for two different filling ratios. The FEM simulation results of the 3D waveguides agree with the analytical results based on the 2D slab waveguides. Effective indices beyond 9 are obtained. As the gap size is less than 5 nm, the coupling between two metamaterial waveguides is getting stronger so that the effective indices will increase rapidly. The mode area is defined as $A_m \equiv \frac{\iint W(x,y)dxdy}{\max[W(x,y)]}$, where $W(x,y)$ is the electromagnetic energy density, and the optical propagation length is calculated from $L_m \equiv 1/2\text{Im}(\beta)$. For the filling ratio of $f_m = 0.4$, the subwavelength mode area $A_m$ is reduced from $3.5 \times 10^{-3} \lambda_0^2$ at $g = 10$ nm, to $1.4 \times 10^{-3} \lambda_0^2$ at $g = 5$ nm, while the propagation length $L_m$ is decreased as well, from 509 nm at $g = 10$ nm, to 487 nm at $g = 5$ nm.

According to Eq. (3), metamaterial slot waveguides with large effective indices $n_{\text{eff},z}$ have high field decay rates in air $\gamma$, indicating that the optical field enhancement will strongly depend on the gap sizes $g$ [as shown in Fig. 2(b)]. On the other hand, the optical mode profile of the metamaterial slot waveguide is distinguished from the case in a conventional dielectric waveguide, due to the extreme anisotropy of hyperbolic metamaterials. For an uncoupled metamaterial waveguide, where $\varphi$ becomes zero in the second formula of Eq. (4), a negative $\varepsilon_z$ in hyperbolic metamaterials implies that $k_y h/2 \in [\pi/2, \pi]$, so that there will be zero values for the $E_y$ field along the $y$ direction inside an individual waveguide. This property will remain for the two coupled metamaterial waveguides, as noticed in Fig. 2(b), where positive $E_y$ in the slot region is accompanied with negative $E_y$ inside both waveguides. In order to take into account this optical mode property of individual metamaterial waveguide and the mode coupling between the two waveguides, here the electric field enhancement factor $\eta$ is defined as the ratio of the electric fields at the two boundaries of each waveguide, i.e., $\eta \equiv \dfrac{E_y\big|_{|y|=\left(\frac{g}{2}\right)^-}}{E_y\big|_{|y|=\left(\frac{g}{2}+h\right)^+}}$, which is unity for an uncoupled waveguide. As the gap size is getting smaller, $E_y$ field is enhanced greatly and $\eta$ increases, as shown in Fig. 3(b). An analytical expression for $\eta$ can be derived using the above theoretical analysis on the 2D coupled slab waveguides,

$$\eta = \sqrt{\frac{\cos^2\left(-k_y \frac{h}{2} + \varphi\right)}{\cos^2\left(k_y \frac{h}{2} + \varphi\right)}} = \sqrt{\frac{1 + \left(\frac{\gamma \varepsilon_z}{k_y}\right)^2}{1 + \left(\frac{\gamma \varepsilon_z}{k_y}\right)^2 \tanh^2(\gamma \frac{g}{2})}} \approx \sqrt{\frac{1 + \varepsilon_y |\varepsilon_z|}{1 + \varepsilon_y |\varepsilon_z| \tanh^2(\gamma \frac{g}{2})}} \qquad (5)$$

where the approximation $\left(\dfrac{\gamma\varepsilon_z}{k_y}\right)^2 \approx \varepsilon_y|\varepsilon_z|$ is obtained by substitution of Eq. (3) under the condition of $k_y \gg k_0$ and $\gamma \gg k_0$. It is clear from Eq. (5) that electric field enhancement factor $\eta$ depends on the gap size $g$, the field decay rate in air $\gamma$, and metamaterial permittivity tensor. For a higher filling ratio $f_m$, both $\gamma$ and $\varepsilon_y|\varepsilon_z|$ have larger values so that $\eta$ will exhibit a larger growth rate as the gap size $g$ decreases, as illustrated in Fig. 3(b) for $f_m = 0.4$ and $f_m = 0.7$. Both $\varepsilon_y$ and $\varepsilon_z$ of the permittivity tensor play an important role in the electric field enhancement. Although $\varepsilon_y$ directly dominates the electric field discontinuity at the metamaterial slot waveguide interfaces, $\varepsilon_z$ will also affects the magnitude of electric field at the slot interfaces and therefore the field enhancement.

As a result of the electric field enhancement, optical power flow $P_{slot}$ and averaged optical intensity $I_{slot} = P_{slot}/wg$ inside the slot region can also be enhanced dramatically. Fig. 3(c) and (d) show the calculated $P_{slot}$ and $I_{slot}$ (normalized to the incident optical power flow) as a function of the gap size $g$ for different filling ratios. The fraction of optical power flow inside the slot region grows up as gap size $g$ decreases, due to the enhanced electric field. For $f_m = 0.4$, $P_{slot}$ will reach a maximum when $g$ is very small, since the field enhancement cannot compensate the shrink of the slot area $w \cdot g$ anymore. On the other hand, the average optical intensity $I_{slot}$ keeps increasing as the gap size $g$ becomes smaller. As light can be strongly compressed and tightly confined in the nanoscale slot waveguide, optical intensity up to 2000 $\mu m^{-2}$ within the slot region is achieved, which is over 25 times stronger than that in a silicon slot waveguide (with a maximum of 80 $\mu m^{-2}$ [1]). The analytical expression for optical power flow $P_{slot}$ and averaged optical intensity $I_{slot}$ can be approximately derived as well,

$$P_{slot} \approx \frac{r_g}{1+r_g}$$
$$I_{slot} = \frac{P_{slot}}{wg} \approx \frac{1}{wg}\frac{r_g}{1+r_g}$$
(6)

where $r_g \approx \dfrac{\varepsilon_y}{h} \dfrac{\dfrac{g}{2}\left[1+\dfrac{\sinh(\gamma g)}{\gamma g}\right]}{1+\varepsilon_y|\varepsilon_z|\sinh^2(\gamma \dfrac{g}{2})}$ is the ratio of optical power flow within the slot region to that inside the two metamaterial waveguides. In the derivation of Eq. (6), the power flow in the exterior surrounding air is neglected. The analytical results can give a clear explanation on the gap size dependence of optical power flow and optical intensity inside the slot region, which are related to the product of $\gamma g$ with hyperbolic-sine-functions.

In reality, the slot waveguide made of metal-dielectric multilayer metamaterials can be constructed to achieve the optical field enhancement predicted by the previous analysis from the effective medium theory. Fig. 1(a) shows the waveguide based on silver-germanium multilayer structures with a period of 10 nm. Fig. 4 shows the electric field $E_y$ profiles along the symmetry line $x = 0$ for $g = 5$ nm with two different filling ratios, calculated from the FEM simulation. In the slot region and the exterior surrounding air, the $E_y$ profiles calculated from multilayer structures perfectly overlap with the EMT prediction. Inside the multilayer metamaterial waveguide cores, $E_y$ field will jump around, due to the coupling of gap plasmons between metal-dielectric interfaces. The spatially homogenized $E_y$, however, is still consistent with the EMT calculation. As shown in Fig. 4(b), multilayer structures with a larger filling ratio give a stronger $E_y$ field on the metal-dielectric interfaces, due to a more tightly confined gap plasmon.

In conclusion, we have presented a new type of slot waveguides made of hyperbolic metamaterials for achieving giant electrical field enhancement in the slot region. A newly defined electric field enhancement factor is used to take the metamaterial waveguide mode profiles and mode coupling properties into account. It is revealed both numerically and analytically that the optical field enhancement critically depends on gap sizes and metamaterial permittivity tensors. Optical intensity up to 2000 $\mu m^{-2}$ is achieved inside the slot region, which is over 25 times larger than that in a conventional silicon slot waveguide. The field enhancement effect is also verified with realistic metal-dielectric multilayer structures. The demonstrated metamaterial slot waveguides with extremely tight photon confinement will open up opportunities for many applications in enhanced light-matter interactions.

**Acknowledgements**

This work was partially supported by the Department of Mechanical and Aerospace Engineering, the Materials Research Center (MRC), the Intelligent Systems Center (ISC), and the Energy Research and Development Center (ERDC) at Missouri S&T, the University of Missouri Research Board, the Ralph E. Powe Junior Faculty Enhancement Award of the Oak Ridge Associated Universities (ORAU), and the National Natural Science Foundation of China (61178062 and 60990322).

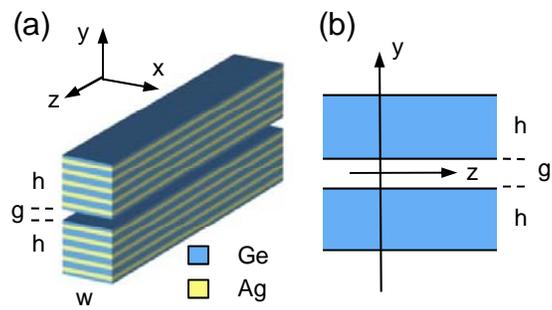

**Figure 1.** Schematic of the nanoscale slot waveguides of hyperbolic metamaterials. (a) 3D silver-germanium multilayer structures, and (b) approximated 2D slab structures for theoretical analysis.

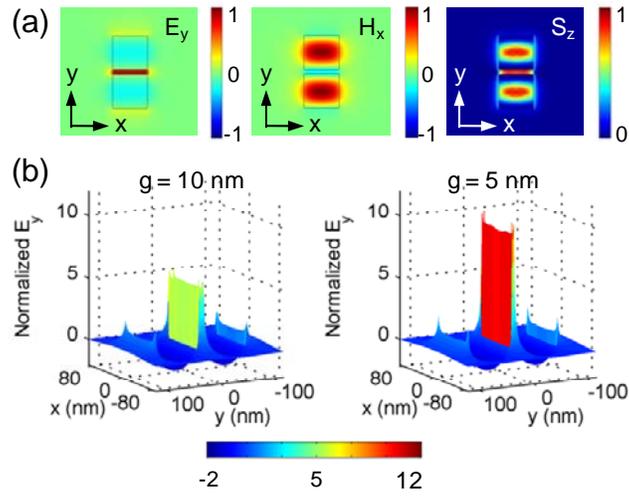

**Figure 2.** (a) Mode profiles of the metamaterial slot waveguide with $g = 10$ nm at $\lambda_0 = 1.55$ μm, where the filling ratio of silver $f_m = 0.4$. (b) 3D surface plots of the $E_y$ field distributions (normalized to $E_y|_{x=0, y=|h+\frac{g}{2}|^+}$) for slot waveguides with $g = 10$ nm and $g = 5$ nm.

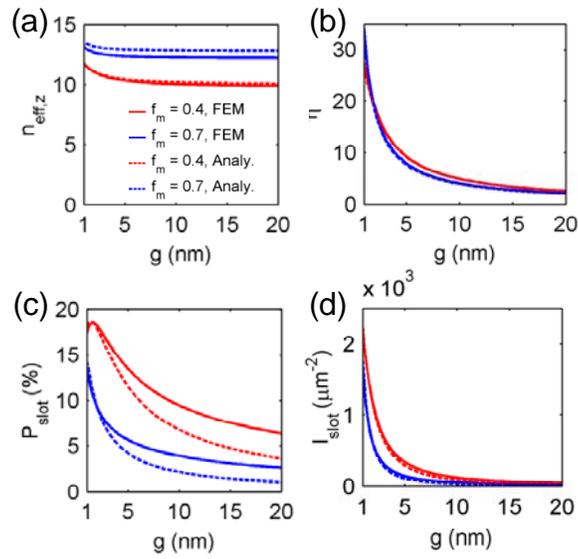

**Figure 3.** (a) Effective refractive indices along the propagation direction $n_{\text{eff},z}$, (b) electric field enhancement factor $\eta$, (c) optical power flow in the slot region $P_{\text{slot}}$, and (d) optical intensity in the slot region $I_{\text{slot}}$ as a function of the gap size $g$ for two different filling ratios of $f_m = 0.4$ and $f_m = 0.7$. Results from both FEM simulation and theoretical analysis are plotted for comparison.

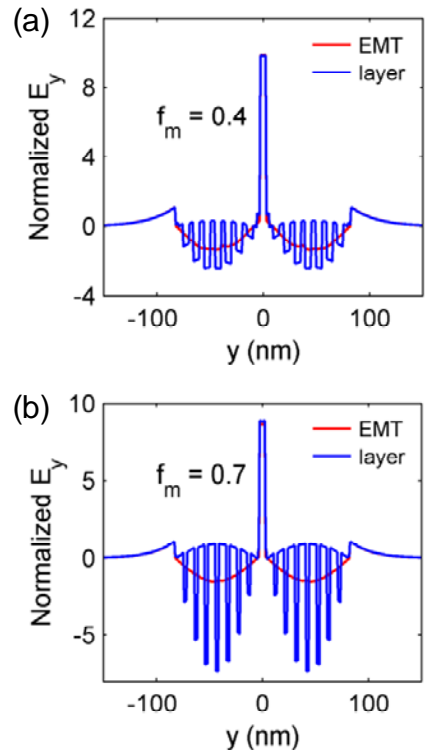

**Figure 4.** Comparison between the $E_y$ field profiles calculated from the effective medium theory (EMT) and realistic metal-dielectric multilayer structures (layer). The mode profiles along $x = 0$ for slot waveguides with $g = 5$ nm are shown in (a) for $f_m = 0.4$ and (b) for $f_m = 0.7$, respectively.